\documentclass{aastex631}

\newcommand{\half}{{\frac{1}{2}}}
\newcommand{\onqu}{{\frac{1}{4}}}
\newcommand{\thqu}{{\frac{3}{4}}}

\shorttitle{A Co-Scaling Grid for Athena++: II Magnetohydrodynamics}
\shortauthors{Heitsch \& Habegger}

\graphicspath{{./}{figures/}}

\begin{document}
\bibliographystyle{aasjournal}

\title{A Co-Scaling Grid for Athena++ II:  Magnetohydrodynamics}

\author[0000-0002-4775-039X]{Fabian Heitsch}
\affiliation{University of North Carolina, Dept. of Physics and Astronomy, Chapel Hill, NC 27599, US}

\author[0000-0003-4776-940X]{Roark Habegger}
\affiliation{University of Wisconsin-Madison Astronomy Department, Madison, WI 53703, US}

\newcommand{\rh}[1]{ {\color{red} #1} }

\newcommand{\revise}[1]{\textbf{#1}}

\begin{abstract}
We extend the co-scaling formalism of Habegger \& Heitsch (2021) implemented in Athena++ to magneto-hydrodynamics. The formalism relies on flow symmetries in astrophysical problems involving expansion, contraction, and center-of-mass motion. The formalism is fully consistent with the upwind constrained transport method implemented in Athena++ and is accurate to 2nd order in space. Applying our implementation to standard magneto-hydrodynamic test cases leads to improved results and higher efficiency, compared to the fixed-grid solutions.
\end{abstract}


\section{Introduction} \label{sec:Intro}

While there is no doubt that interstellar gas is magnetized, the role of magnetic fields for the evolution especially of the denser gas phases down to star formation regions is hotly debated (for a recent summary, see \citealp{Pattle2022}). Magnetic fields can affect the evolution of blast waves because they break the radial symmetry of the flow, leading to an effective ``sweep-up" of material toward the equatorial region \citep{Ferriere1991}. The evolution of magnetic fields around expanding bubbles may also be of interest for understanding the formation of molecular clouds and stars in a triggered star formation scenario, since the observed field strengths tend to be larger than the expected average field strength in the interstellar medium \citep{Bracco2020}.

Covering large spatial ranges in numerical simulations of e.g. supernova or kilonova remnants has been addressed in a variety of ways. General-purpose approaches which do not exploit a given problem's symmetries include Lagrangian methods like smoothed particle hydrodynamics \citep{Monaghan1992} or adaptive mesh refinement techniques for Eulerian codes \citep[e.g.][]{2000Fryxell,2006Fromang, Krumholz2007,Klein2017,Stone2020}. Additionally, there are moving-mesh codes, which solve flux-conservative problems on meshes that move with the fluid in a Lagrangian fashion \citep{Hopkins2015,2010Springel}. 
While conceptual limitations of early implementations of magnetic fields in smoothed particle hydrodynamics \citep{Brandenburg2010,Price2010,PriceFederrath2010} have been overcome \citep{StasyszynElstner2015}, concerns regarding fulfillment of the divergence constraint remain \citep{WissingShen2020}.

A conceptually simpler alternative to the above is to exploit possible symmetries in an astrophysical problem. This exploitation can be more efficient \citep{Roepke2005} while also preserving uniformity of dissipative properties across the grid. More recently, \citet[][see also \citealp{RobertsonGoldreich2012} along similar lines for an implementation to study adiabatically driven turbulence]{SunBai2023} presented a co-moving domain implementation for MHD employing an expanding coordinate frame similar to cosmological simulations \citep[e.g.][]{Bryanetal2014}. A specific implementation of an accelerated expanding box for MHD to model turbulence in the solar wind is discussed by \citet{TeneraniVelli2017}. \cite{Xuetal2024} extend the local shearing box model to a collapsing or expanding sphere for hydrodynamics, following a local patch and modifying the pressure and energy terms such that signal speeds (and thus wavelengths) stay constant within the domain.

In a previous study (\citealp{HabeggerHeitsch2021}, HH21), we implemented a co-scaling mesh in the Eulerian grid code Athena++ \citep{Stone2020}\footnote{\href{https://github.com/roarkhabegger/athena-TimeDependentGrid}{https://github.com/roarkhabegger/athena-TimeDependentGrid}}. Instead of modifying the underlying equations by scaling factors as in cosmological codes or by varying the signal speeds \citep{Xuetal2024}, we accounted for the domain expansion by adding fluxes associated with the cell-wall motion. In that sense, the implementation is ``minimally invasive" and can be fully decoupled from the standard version of Athena++.
The grid can be co-moving or rescaled, retaining the initial cell aspect ratio. The grid evolution is integrated at the same time order as the fluid variables. The time dependence of the grid scaling is defined by a user-specified function. The co-scaling grid can be combined with the adaptive mesh capabilities of Athena++. 

Here, we extend the co-scaling grid formalism to full three-dimensional magnetohydrodynamics in Cartesian and spherical-polar coordinates\footnote{\href{https://github.com/fheitsch/athena-public-version/tree/expgrid}{https://github.com/fheitsch/athena-public-version/tree/expgrid}}.  One-dimensional shock tests are improved using the co-scaling grid. Two- and three-dimensional tests illustrate the consistency of the co-scaling grid with fixed grid simulations. We find the method reproduces standard test cases at high accuracy and reduces spurious $\nabla\cdot\textbf{B}$ in higher dimensions.

%%%%%%%%%%%%%%%%%%%%%%%%%%%%%
%%%%%  FORMALISM       %%%%%%
%%%%%%%%%%%%%%%%%%%%%%%%%%%%%

\section{Formalism} \label{sec:Form}
As summarized in HH21, Eulerian, ideal magnetohydrodynamics solve the  conservation laws
\begin{equation}
   \frac{\partial \textbf{U}^T(\vec{x},t)}{\partial t}
   = -\vec{\nabla}^T\cdot \overline{\textbf{$\Gamma$}} (\vec{x},t) .
   \label{eqn:PDE}
\end{equation}
The row vector $\textbf{U}^T$ contains the conservative variables.
The matrix $\overline{\textbf{$\Gamma$}}$ has columns with the flux of each conservative quantity. These fluxes have rows corresponding to the various coordinate directions ($\hat{x}_1$, $\hat{x}_2$, and $\hat{x}_3$) \citep{Stone2008}. The length of $\textbf{U}^T$ depends on the physics of the problem.  For ideal MHD, $\textbf{U}^T$ has $8$ components and the matrix $\overline{\textbf{$\Gamma$}}$ has $3$ rows and $8$ columns.  Altogether, the right hand side is the flux divergence. For a Cartesian grid, the matrix $\overline{\textbf{$\Gamma$}}$ has the form
\begin{equation}
  \overline{\textbf{$\Gamma$}} = \textbf{F}^T\hat{x} + \textbf{G}^T\hat{y} + \textbf{H}^T\hat{z}
  \label{eqn:FluxMatrix}
\end{equation}
where each boldfaced vector of conservative variables is the flux of those quantities in the given direction.

By integrating Equation\,\ref{eqn:PDE} over a discrete volume $\Delta V$, the differential equation becomes an integro-differential equation. For static grids, this equation can be rewritten as an ordinary differential equation for the conservative variables $\textbf{U}$ of each cell, indexed by $(i,j,k)$:
\begin{eqnarray}
    \frac{d}{dt} \textbf{U}_{i,j,k} = 
        &-& \frac{1}{\Delta x_{i}}\left( 
        \textbf{F}_{i+\frac{1}{2},j,k} 
        -  \textbf{F}_{i-\frac{1}{2},j,k}\right)\nonumber \\
        &-& \frac{1}{\Delta y_{j}}\left( 
        \textbf{G}_{i,j+\frac{1}{2},k}
        -  \textbf{G}_{i,j-\frac{1}{2},k} \right)\nonumber \\
        &-& \frac{1}{\Delta z_{k}}\left( 
        \textbf{H}_{i,j,k+\frac{1}{2}}
        -  \textbf{H}_{i,j,k-\frac{1}{2}}\right)
    \label{eqn:StaticUpdate}
\end{eqnarray}
where the conservative variables $\textbf{U}$ are averaged over the cell volume and the flux vectors $\textbf{F}$, $\textbf{G}$, $\textbf{H}$ are averaged over a cell wall (see \citealp{Stone2020,Felker2018}). A more detailed derivation of Eqn.\,\ref{eqn:StaticUpdate} can be found in Appendix~A of \citet{HabeggerHeitsch2021}.

In principle, the magnetic terms can be implemented - as the hydrodynamical ones - as cell-centered quantities, via 
\begin{eqnarray}
\frac{\partial}{\partial t}\textbf{B}&=& -\vec{\nabla}\times \textbf{E}\nonumber\\
  &=&\vec{\nabla}\times(\textbf{v}\times\textbf{B}),
\end{eqnarray}
with the electric field $\textbf{E}\equiv -\textbf{v}\times\textbf{B}$. Yet, cell-centered MHD implementations require additional steps \citep{Dedner2002,WangAbel2009} to preserve the divergence-free constraint of the magnetic field. A natural way to keep $\vec{\nabla}\cdot\textbf{B}\equiv 0$ is to interpret the magnetic field as an integral over the cell area instead of the volume \citep{GardinerStone2005}. This method results in a constrained transport formulation that preserves the divergence-free constraint by construction to machine accuracy \citep{EvansHawley1988}:
\begin{eqnarray}
\frac{\partial}{\partial_t}\Phi&\equiv&\frac{\partial}{\partial_t}\int_{\Sigma}\textbf{B}\cdot d\textbf{A}\nonumber\\
  &=& \int_\Sigma \vec{\nabla}\times(\textbf{v}\times\textbf{B})\cdot d\textbf{A}\nonumber\\
  &=& \oint_{\partial \Sigma} (\textbf{v}\times\textbf{B})\cdot d\textbf{s}.\label{eqn:MHDflux1}
\end{eqnarray}
The last integral is not over the area surface $\Sigma$ but over the closed area boundary $\partial \Sigma$. Such methods usually are implemented using a staggered mesh to allow for placement of the field variables on the cell faces/edges \citep{StoneNorman1992,BalsaraSpicer1999}.

As in the hydrodynamic case (HH21), integration and time derivatives only commute for a static grid. The justification for that step is the Reynolds Transport Theorem for a quantity $f$ over a volume $V$ and boundary $B$ moving at velocity $\vec{w}$,
\begin{equation}
    \frac{d}{dt}\int_V dV \, f
    = \int_V dV \, \frac{\partial f}{\partial t} 
    + \int_{B} dA \, (\vec{w} \cdot \hat{n}) f .
    \label{eqn:RTT}
\end{equation}
For the magnetic field, the corresponding integral reads \citep{Blackman2013} 
\begin{equation}
\frac{d}{dt}\int_\Sigma \textbf{B}\cdot d\textbf{A}=\int_\Sigma\frac{\partial\textbf{B}}{\partial t}\cdot d\textbf{A}
-\oint_{\partial\Sigma}(\textbf{w}\times\textbf{B})\cdot d\textbf{s}.\label{eqn:totalMHDcoscaling}
\end{equation}
Combining Equations~(\ref{eqn:MHDflux1}) and (\ref{eqn:totalMHDcoscaling}) yields the MHD induction equation in terms of the magnetic flux including time-dependent areas,
\begin{equation}
\frac{d}{dt}\int_{\Sigma}\textbf{B}\cdot d\textbf{A}
  = \oint_{\partial \Sigma} (\textbf{v}\times\textbf{B})\cdot d\textbf{s}
  -\oint_{\partial \Sigma} (\textbf{w}\times\textbf{B})\cdot d\textbf{s}
  \label{eqn:MHDflux2}
\end{equation}
Equation~\ref{eqn:MHDflux2} can also be derived by subtracting the wall velocity $\textbf{w}$ from the bulk velocity $\textbf{v}$.  Note that the curve $\partial\Sigma$ is not a material curve, i.e. it is not tied to the gas, but it  describes the change of the control area. 

In addition to the wall fluxes, the volume-averaged, hydrodynamical quantities require a second correction (see HH21, Equation~12) accounting for the volume change. Analogously, all field components need to be rescaled by the updated area, to keep the update consistent with the conservative formulation of the equations. For example, the third field component $B^3$ needs to be corrected as 
\begin{equation}
    B^3_{i,j,k}(t^{n+1}) = \frac{A^3_{i,j,k}(t^n)}{A^3_{i,j,k}(t^{n+1})}
    \Bigg[ B^3_{i,j,k}(t^{n})%\nonumber \\ 
    + 
    \int_{t^n}^{t^{n+1}} dt \left[ \frac{d}{dt} B^3_{i,j,k} \right] \Bigg].
    \label{eqn:MHDTimeInt}
\end{equation}
The surface area $A^3_{i,j,k}(t)$ is perpendicular to $B^3$ of the $(i,j,k)$ cell at time $t$. For a uniform Cartesian grid, $A^3_{i,j,k}=\Delta x_i \Delta y_j$. We use this opportunity to point out that Equations~8 and 9 of HH21 contain an incorrect normalization of the integral by $t_{n+1}-t_n$. 
Since each cell changes by the same factor along all coordinate axes, Equation~\ref{eqn:MHDTimeInt} implies that the field at timestep $n+1$ is kept divergence-free, $\vec{\nabla}\cdot\textbf{B}^{n+1}=0$, if the field at timestep $n$ was divergence-free.

Thus, the time-dependent grid requires two corrections to the magnetic flux. The first is to include 
the magnetic flux change arising from the change in the length of the area boundary (Equation~\ref{eqn:MHDflux2}). The second applies the change in cell area,  scaling the conserved magnetic flux. (Equation~\ref{eqn:MHDTimeInt}).

\noindent

%%%%%%%%%%%%%%%%%%%%%%%%%%%%%
%%%%%  Implementation  %%%%%%
%%%%%%%%%%%%%%%%%%%%%%%%%%%%%

\section{Implementation}\label{sec:Impl}
Athena++ solves Equation\,\ref{eqn:PDE} over a static grid \citep{Stone2008,Stone2020}. A co-scaling grid requires the integration of the grid's motion over time, in addition to the integration of the physical variables. After this grid integration, we add corrections to the physical variables in the form of boundary source terms to the induction Equation (Sec.~\ref{ss:boundarysource}), and area scaling (Sec.~\ref{ss:areascaling}). Finally, all coordinate variables need to be updated throughout the full mesh hierarchy, including derived quantities such as cell volumes and areas, and reconstruction coefficients. This requires changes to the task list implemented in Athena++. The time integration of the grid, the update of the coordinates, and the changes to the task list have been described in HH21. Here, we report on the details of the implementation of Equations~\ref{eqn:MHDflux2}~\&~\ref{eqn:MHDTimeInt}. 

\subsection{Source Terms}\label{ss:boundarysource}

\begin{figure*}[t!]
  \centering
 \includegraphics[width=0.9\textwidth]{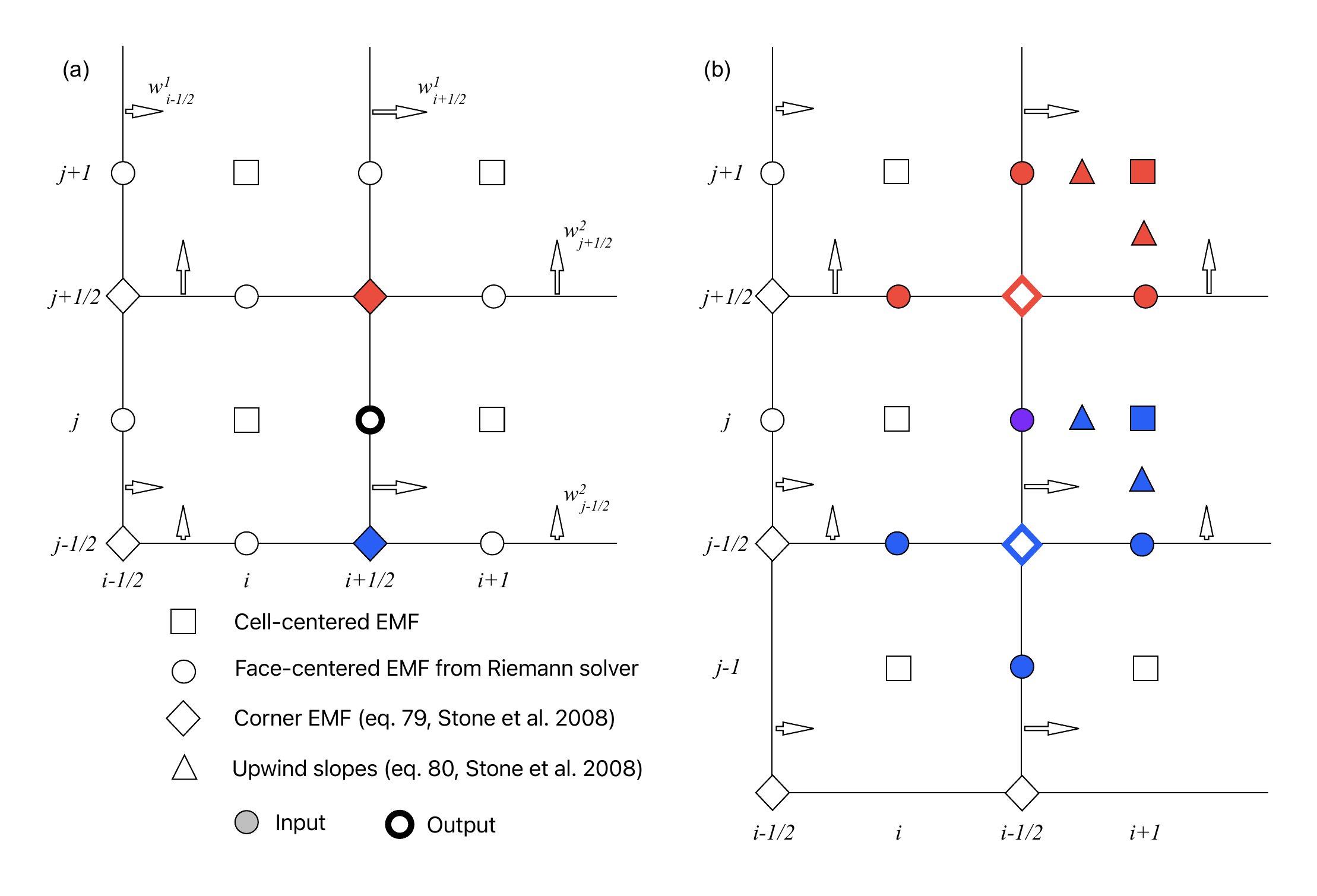}
  \caption{\label{fig:geometry}Update sequence and stencil extent for a uniformly expanding Cartesian grid, assuming that the second field component $B^2=0$. Wall velocities at constant $i$ and $j$ are constant, as indicated by the arrows. Filled symbols denote input values, thick symbols stand for output. (a) Positions used for the induction equation, Equation~\ref{eqn:db1dt}. (b) Positions used for the upper ($(i+1/2,j+1/2)$, red) and lower ($(i+1/2,j-1/2)$, blue) corner EMF (Equation~\ref{eqn:e3full}) contributing to face-centered $B^1_{i+1/2,j}$ shown in (a). The upwind conditions (e.g. Equation 30, 33 of \citealp{Felker2018}) are indicated for the assumed wall motion (left to right, bottom to top).}
\end{figure*}

Athena++ solves the induction equation in two steps. In a first step, the electromotive force (EMF) 
\begin{equation}
  \textbf{E}=-\textbf{v}\times\textbf{B}
\end{equation}
is calculated and integrated out to the cell corners, and in the second step, the EMF at the corners is used to update the magnetic flux,
\begin{equation}
  \partial_t \Phi = -\oint_{\partial\Sigma} \textbf{E}\cdot d\textbf{s}. 
\end{equation}
The expanding grid causes a second integral (Equation~\ref{eqn:MHDflux2}) due to the apparent EMF caused by the wall velocity $\textbf{w}$.

For the interpolation of the EMF to the cell corners, Athena++ requires the velocities and fields at the cell centers and at the cell walls. We follow the implementation of the 2nd-order accurate, upwind constrained transport method introduced by \citet{GardinerStone2005} in the formulation of \citet{Stone2008,Felker2018}. We describe the approach for the $B^1$ component of the magnetic field in the $(x_1,x_2)$-plane. Figure~\ref{fig:geometry} shows the location of all quantities. 

The time evolution of the face-centered $B^1$ component is given by
\begin{equation}
  \frac{d}{dt}B^1_{i-\half,j}=-\frac{1}{\Delta x_2}\left(\langle{\cal{E}}^3\rangle_{i-\half,j+\half}-\langle{\cal{E}}^3\rangle_{i-\half,j-\half}\right),\label{eqn:db1dt}
\end{equation}
with the line-averaged, cell-corner EMF
\begin{eqnarray}
  \langle{\cal{E}}^3\rangle&=&\frac{1}{4}\left({\cal{E}}^3_{i,j-\half}+{\cal{E}}^3_{i-1,j-\half}+{\cal{E}}^3_{i-\half,j}+{\cal{E}}^3_{i-\half,j-1}\right)\nonumber\\
  &+&\frac{\Delta x_2}{8}\left(\left(\frac{\partial{\cal{E}}^3}{\partial x_2}\right)_{i-\half,j-\thqu}-\left(\frac{\partial{\cal{E}}^3}{\partial x_2}\right)_{i-\half,j-\onqu}\right)\nonumber\\
  &+&\frac{\Delta x_1}{8}\left(\left(\frac{\partial{\cal{E}}^3}{\partial x_1}\right)_{i-\thqu,j-\half}-\left(\frac{\partial{\cal{E}}^3}{\partial x_1}\right)_{i-\onqu,j-\half}\right).
  \label{eqn:e3full}
\end{eqnarray}
The first four terms are the EMFs along the four sides constituting the line integral around the cell, calculated by the Riemann solver. The first two terms arise from the fluxes along $x_2$, and the remaining two terms come from the fluxes along $x_1$. For the fluxes along $x_1$,  
\begin{equation}
  {\cal{E}}^3_{i-\half,j}=v^1_{i-\half,j}B^2_{i-\half,j}-v^2_{i-\half,j}B^1_{i-\half,j},
\end{equation}
where $v^1_{i-\half,j}$ is derived from the momentum flux and $B^1_{i-\half,j}$ is the (inactive) face-centered field component. The other two quantities are reconstructed values at the face center. That completes the standard implementation. To account for the moving wall, we add the apparent EMF 
\begin{equation}
{\mathfrak{E}}^3_{i-\half,j}=-w^1_{i-\half,j}B^2_{i-\half,j}+w^2_{i-\half,j}B^1_{i-\half,j}\label{e:apparentemf}
\end{equation}
at the end of the flux computation in the Riemann solver. The field components are the same as above. The wall velocity along the update direction (here $w^1$) is located at the cell wall, as provided by our previous implementation (HH21). The cross-velocity $w^2_{i-\half,j}$ is, because of the rectilinear grid expansion, identical to the cell-centered $w^2_{i,j}$, and therefore can be easily generated at the appropriate locations together with the face-centered values of $w^2$ during the grid expansion step. This cell-centered wall velocity will also be necessary for the remaining terms in Equation~\ref{eqn:e3full}. 

The remaining terms of Equation~\ref{eqn:e3full} involving derivatives require two modifications in the code. First, to maintain the upwind condition, the calculation of the interpolation weights needs to be modified to include the wall velocity. Using the example of Eq.~30 in \citet{Felker2018} for the expression for upwinding the $\partial_2$ derivatives to the $x_1$ faces, the condition changes from $v^1_{i-1/2,j}\lessgtr 0$ to $v^1_{i-\half,j}-w^1_{i-\half,j} \lessgtr 0$.
Second, the derivatives located at e.g. positions $i-\onqu$ in Equation~\ref{eqn:e3full} require the cell-centered EMF, for which the (cell-centered) velocities need to be modified by subtracting the corresponding cell-centered wall velocities (see above). Since the face-centered EMF terms already include the wall-contribution $\mathfrak{E}^3$ via the Riemann solver, the interpolation to determine the line-averaged EMF is now complete. 

\subsection{Area Scaling}\label{ss:areascaling}
Since the fully conservative formulation of the induction equation refers to the magnetic flux (Equation~\ref{eqn:MHDflux1}), and not the magnetic field, in a last step the change in cell area needs to be taken into account, similar to the volume change for hydrodynamics. To keep consistent with the integrator structure, the area needs to be calculated ahead of the coordinate update (here for the area element $A_3=\Delta x_1\Delta x_2$) as 
\begin{eqnarray}
  A_3^{n+1}&=& A_3^n + \frac{d}{dt}A^n\Delta t\nonumber\\
         &=& A_3^n + \left(\frac{\partial A_3}{\partial \Delta x_1}\frac{\partial \Delta x_1}{\partial t}+\frac{\partial A_3}{\partial \Delta x_2}\frac{\partial \Delta x_2}{\partial t}\right)\Delta t\nonumber\\
         &=&A_3^n + \left(\Delta x_2\Delta w_1+\Delta x_1 \Delta w_2\right)\Delta t
\end{eqnarray}
where $\Delta w_1=w^1_{i+1/2}-w^1_{i-1/2}$ etc is the difference of the wall velocities. Expressions for spherical-polar coordinates are provided in the appendix.

\subsection{Time Stepping}\label{ss:timestep}
Two limits need to be imposed in addition to the usual CFL timestep restrictions to guarantee stable solutions. First, a coscaling grid volume must not move further than its own size. For expanding or contracting grids, velocities are largest furthest away from the reference point, hence those locations will set the timestep. Second, any wave traveling through a grid element must not travel further than the expected new wall location. This condition can be implemented by modifying the standard CFL condition as
\begin{equation}
  \Delta t \leq c_{\rm CFL}\frac{\Delta x}{|v|+|w|},
\end{equation}
where $v$ is the physical signal speed (e.g. the sum of the bulk velocity and the fast magnetosonic speed), and $w$ is the wall velocity. 

%%%%%%%%%%%%%%%%%%%%%%%%%%%%%
%%%%%  Implementation  %%%%%%
%%%%%%%%%%%%%%%%%%%%%%%%%%%%%
\section{Tests}\label{sec:tests}
We present a series of tests to demonstrate accuracy and performance of the co-scaling MHD grid. All tests were run with the 2nd-order Runge-Kutta integrator native to Athena++, and they used 2nd-order (piece-wise linear) reconstruction in the primitive variables. We implemented and tested the expanding grid for the HLLE, HLLD, and Roe solvers. Here we show results for the HLLD solver. The implementation is total-variation-diminishing (App.~\ref{app:tvdcheck}). For the baseline, static grid comparison, we used the original Athena++ implementation. Our current implementation is limited to 2nd-order spatial accuracy \citep{Felker2018}, but an extension to higher spatial order is possible. 

\subsection{1D Brio-Wu}\label{subsec:briowu}
\begin{figure*}[!t]
    \includegraphics[width=\textwidth]{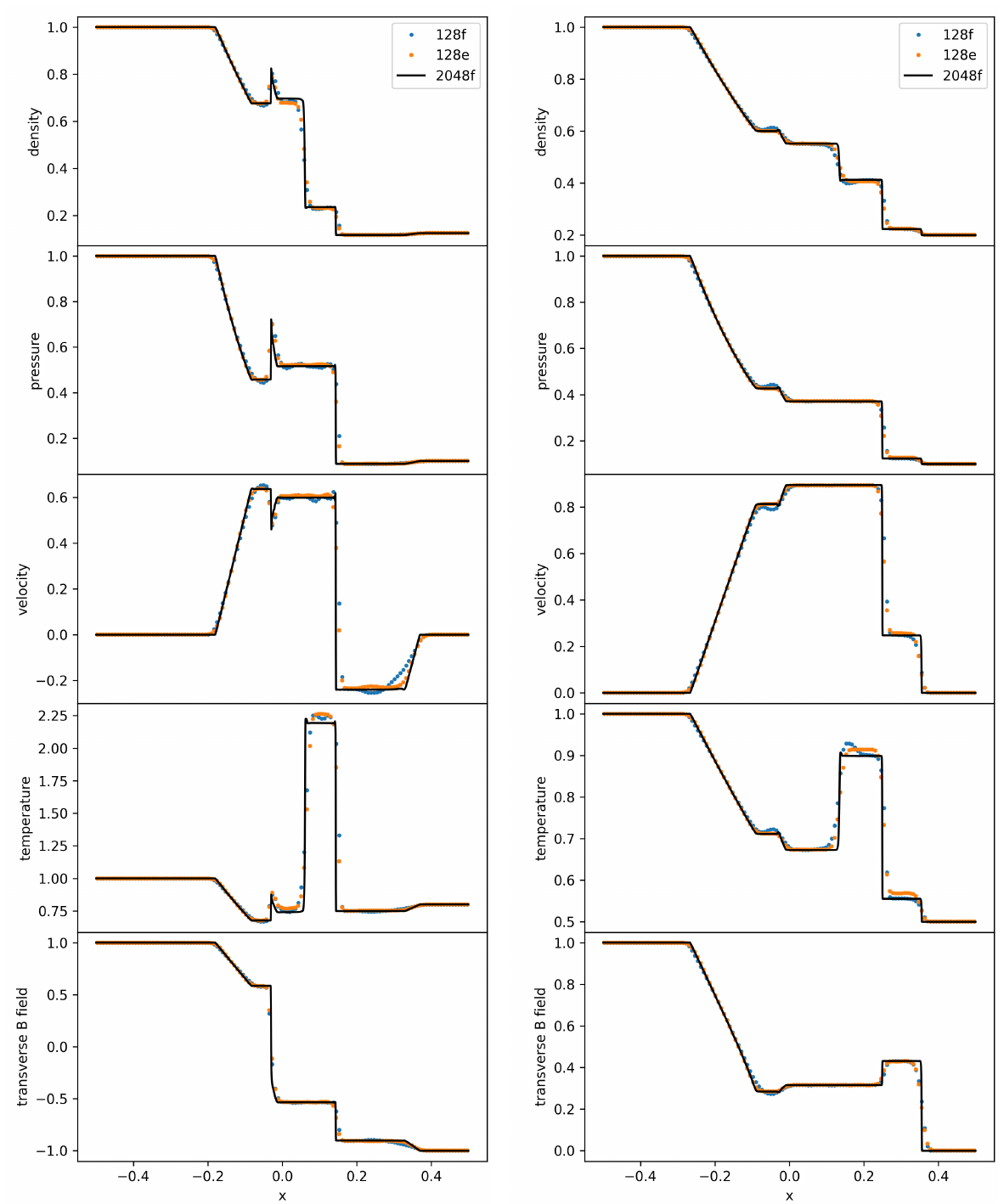}
    \caption{{\em Left:} Density, pressure, velocity, temperature, and transverse field for the \cite{BrioWu1988} test at $t=0.2$. {\em Right:} Same quantities for the switch-on shock test \citep{Balsara1998}. Black lines indicate the reference solution on a fixed grid with $2048$ cells. Blue symbols denote the fixed grid solution at $128$ cells, and orange ones the co-scaling grid solution, also at $128$ cells. Over- and undershoots appearing in the fixed-grid solution are reduced for the co-scaling grid.}
    \label{fig:briowu}
\end{figure*}

The hydrodynamical variables are initialized identically to the Sod shock tube \citep[][Table 1]{Balsara1998}. A constant magnetic field along the $x$-axis, and a transverse field with a discontinuity at $x=0$ is added, with $B_{y,l}=1$ and $B_{y,r}=-1$. \citet{BrioWu1988} discuss the various waves forming. Figure~\ref{fig:briowu} shows the test results at $t=0.2$. The black solid line shows a (fixed grid) reference solution at $2048$ grid points. Blue markers indicate the fixed-grid solution at $128$ points, and orange markers the co-scaling grid solution also at $128$ points. The profiles are nearly identical, except for the temperature, where the oscillations are reduced for the co-scaling grid, and the velocity, where the fast rarefaction wave (right-most gradual slope) is more accurately reproduced by the co-scaling grid.

\subsection{1D Switch-On Shock}
The test demonstrates the formation of a right-going switch-on shock \citep[][Test 1]{Balsara1998}. The left values are $\rho_l=1, p_l=1,B_{x,l}=1,B_{y,l}=1$ and the right values are $\rho_l=0.2, p_l=0.1,B_{x,l}=1,B_{y,l}=0$. The results are shown in the right panel of Fig.~\ref{fig:briowu}. Consistent with the Brio-Wu test, the expanding grid reduces the amplitude of oscillations. Note that when comparing to \citet{Balsara1998}, the expanding grid was run at $128$ points rather than at $800$ points.

\subsection{2D Blast Wave}
We run the standard 2D blast wave test in the implementation of \citet[][see also \citealp{Balsara1998}]{zacharyetal1994}. We initialize a spherical region on a Cartesian grid of extent $[-0.2,0.2]^2$ with a pressure profile 
\begin{equation}
  p(x,y) = p_0+\frac{1}{2}(p_1-p_0)\left(1-\tanh\frac{r-r_0}{w}\right),  
\end{equation}
with the pressures $p_0=0.1,p_1=10^3$, the radius of the sphere $r_0=0.1$ and the width of the $\tanh$-profile $w=10^{-3}$. The density is set to $\rho=1$. The magnetic field is set to a constant value of $B_1/\sqrt{4\pi}=28.21,B_2=0$. The test shown was run on a uniform grid of $200^2$. We ran a comparison model on a fixed grid, initialized with the same parameters on a domain of size $[-0.5,0.5]^2$. Both models are run out to $t=10^{-2}$. For the coscaling grid model, we use a fixed expansion speed of $v_{exp}=30$, so that the physical domain size corresponds to that of the fixed grid model by the end of the simulation. 

\begin{figure*}[t!]
  \includegraphics[width=\textwidth]{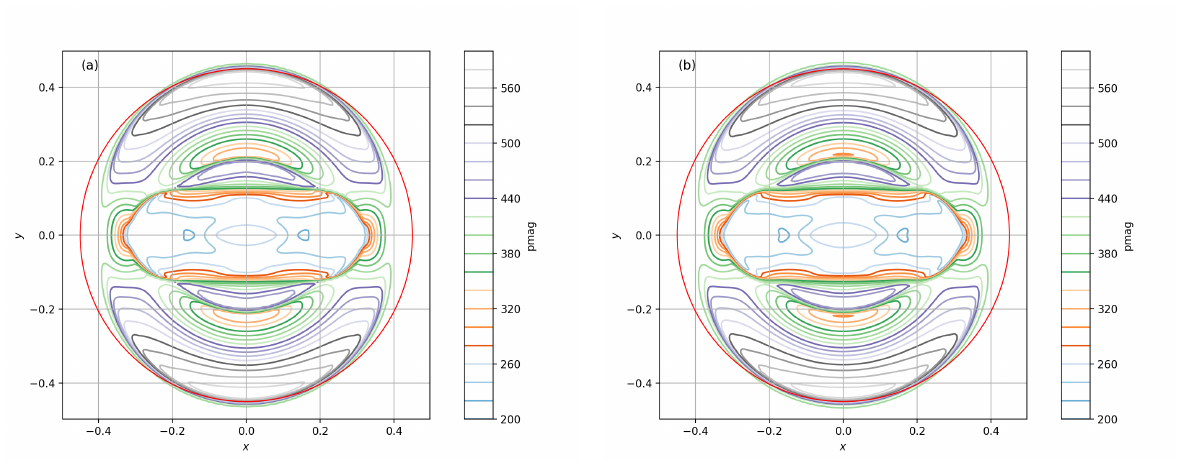}
  \caption{Contour plots of the magnetic pressure for the fixed-grid 2D blast wave (a) and the blast wave run with a co-scaling grid at fixed expansion velocity (b). Contours are spaced equally over $20$ levels. The red circle is shown for orientation.}
  \label{fig:blast2d}
\end{figure*}

\begin{figure}
  \centering
  \includegraphics[width=0.95\linewidth]{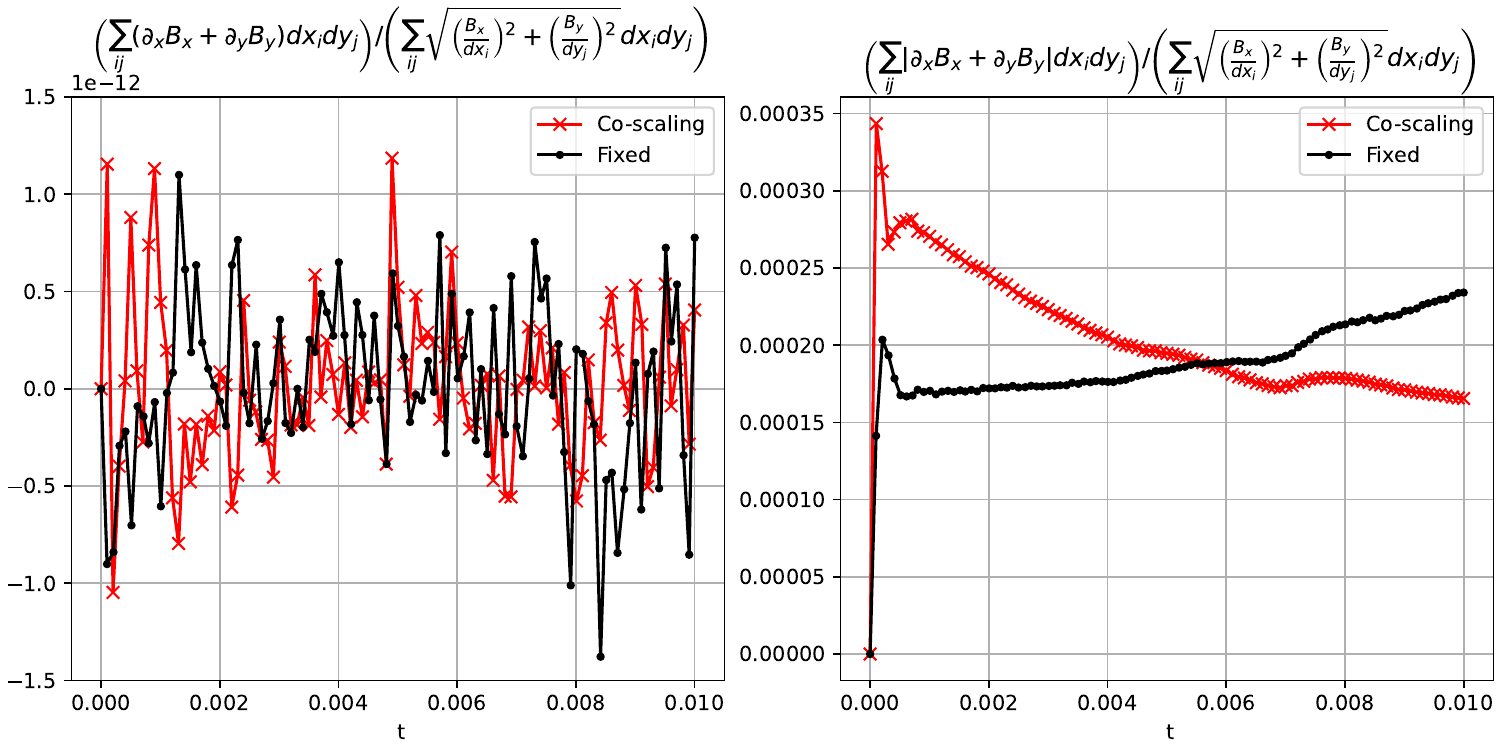}
  \caption{Left: Integrated divergence of the magnetic field for the fixed-grid 2D blast wave (black line with dot markers) and the co-scaling grid 2D blast wave (red line with x markers). The divergence is normalized by the magnetic field strength per cell length integrated over the simulation. This normalization is designed to  leave the vertical axis of this plot in dimensionless units while accounting for the changing grid of the co-scaling simulation. Right: Integration of the absolute value of the divergence of the magnetic field, normalized in the same way as the left plot. Overall, both simulations are good at keeping the magnetic field divergence free over the entire volume. While there is some divergence created at the start of the blast wave's evolution, this is a small error.}
  \label{fig:blast2dDiv}
\end{figure}

Figure~\ref{fig:blast2d} compares the magnetic pressure $B^2/2$ between both runs. The maps are nearly indistinguishable, suggesting that the MHD implementation of the co-scaling grid works to specifications. For additional comparison, we show the evolution of the total divergence of the magnetic field in Figure~\ref{fig:blast2dDiv} for the two simulations. While there should ideally be no divergence, the propagation of the blast leads to some divergence being created in the simulation. The left plot shows the divergence integrated over the entire simulation, whereas the right plot shows the integration of the absolute value of the divergence. Both integrals are normalized by the magnetic field strength per unit length integrated over the simulation, leaving the y-axis of both figures Figure~\ref{fig:blast2dDiv} dimensionless. This normalization is necessary to account for the changing cell size in the co-scaling simulation. The integration of the divergence in both simulations reaches computational noise levels. This consistency between the co-scaling and fixed grid simulations shows there is no additional divergence created during the expansion of the grid. Instead, spurious divergence 
in the co-scaling grid appears only early on, as does in the fixed grid model.  The right plot of Figure~\ref{fig:blast2dDiv} shows this development of divergence: both simulations see a spike in absolute value of the divergence of the magnetic field at the simulation's start. As the co-scaling simulation expands, this total divergence actually decreases, eventually becoming less than in the fixed grid simulation.

\subsection{Colliding 2D Blast Waves}
We initialize two spherical blast waves on a Cartesian grid. Each individual blast wave has the same setup as the test described in the previous section. The simulation domain is extended parallel to the magnetic field with the range of $x_1$ being initially $[-0.6,0.6]$ for the fixed grid and $[-0.3,0.3]$ for the co-scaling grid. The $x_2$ range is unchanged. The blast centers are placed at $x_1 = \pm 0.1$, $x_2=0$. 

\begin{figure*}
  \includegraphics[width=\textwidth]{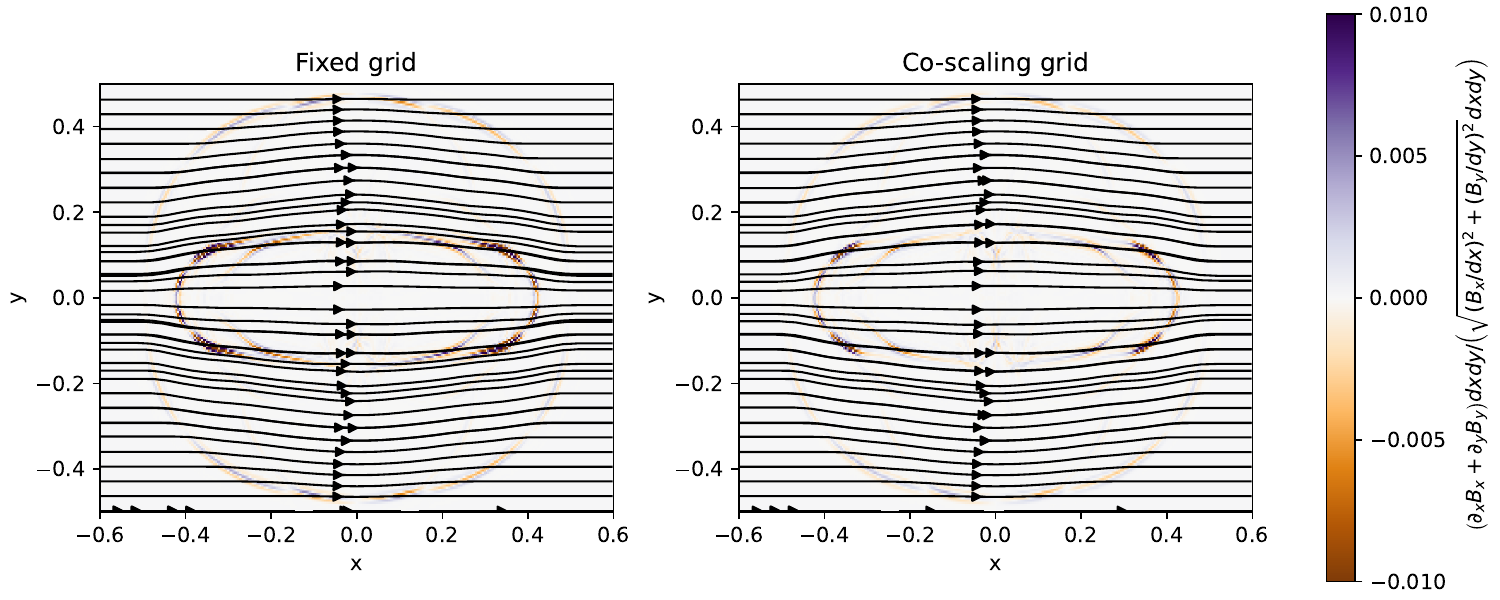}
  \caption{Map of the divergence of the magnetic field in the colliding 2D blast waves simulation. The divergence is calculated at every cell, and then normalized by the magnetic field strength per cell length. As a result, the units of the colorbar are dimensionless. The outer shock matches the 2D blast pictured in Figure~\ref{fig:blast2d}. This simulation has two blast waves, which collide at the $x = 0$ interface, creating a second shock which moves through the simulation. The co-scaling grid has the same structure and fewer locations where neighboring cells flip the sign of $\vec{\nabla} \cdot \vec{B}$. Black streamlines show the magnetic field direction.}
  \label{fig:blast2dcollide}
\end{figure*}

Figure~\ref{fig:blast2dcollide} shows a map of the divergence of the magnetic field for the fixed and co-scaling simulations. We normalize the divergence by the magnetic field strength per cell length. Each simulation has the same structure, as there is no difference in how the blast waves move in the co-scaling and fixed simulations. That agreement also applies to the reflected waves created by the blast waves colliding at the $x_1 = 0$ interface. Local deviations from $\vec{\nabla} \cdot \vec{B} =0$ are reduced in the co-scaling simulation. The deviations only appear because this simulation maps curved blast waves onto a Cartesian grid. Similar to the left hand plot of Figure~\ref{fig:blast2dDiv}, the total divergence integrated over the simulation is near $0$, whereas the integral of the absolute value will be on the order of $10^{-3}$. The vanishing of the total divergence is visible when noting the deviations from $\vec{\nabla} \cdot \vec{B} =0$ are equal and opposite when looking between the left and right sides of each plot. Since the reflections of the colliding shocks are propagating through a low magnetic field strength medium, there is a larger change in the magnetic field strength at those shock fronts. 

\subsection{3D Blast Wave}
To test the full implementation, we run a three-dimensional blast test.
The ambient variables are set to $(\rho_0=1.0,p_0=1.0,v=0.0,B_x=1.0,B_y=1.0,B_z=0)$, resulting in a plasma $\beta=1.0$. The blast is initialized within a sphere of $r_1=0.05$ with $(\rho_1=10.0,p_1=10^4)$. For the co-scaling grid, we start with a domain size of $[-0.1,0.1]^3$ with a resolution of $128^3$, and we stop the evolution at $t=0.15$. 

To follow the blast wave, we implemented a shock tracker. We search for the position of the most negative radial magnetic pressure gradient and determine the speed of the front by finite differencing between the current and previous timestep. For the first $20$ iterations, the grid is stationary, and afterwards, we take the averages of the front velocity over the last $20$ timesteps. This prevents oscillations in the front velocity. The approach works equally well for the hydrodynamic case when replacing the magnetic pressure by the thermal pressure. We note that though the front tracking seems convenient, an analytic prescription derived from fitting an expansion law to a lower-resolution simulation will be more efficient and also leads to numerically more stable results. 

\begin{figure*}
  \includegraphics[width=\textwidth]{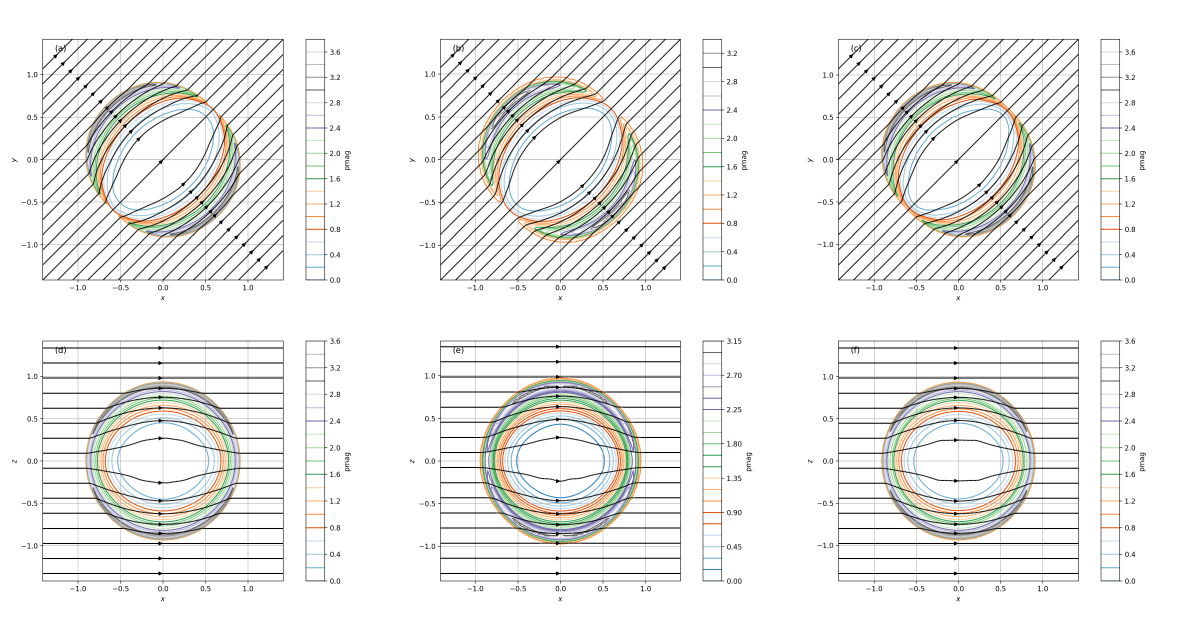}
  \caption{Midplane slices in magnetic pressure $B^2/2$ for the three-dimensional blast test. Shown are slices in the $(x,y)$-plane (top row) and $(x,z)$-plane (bottom row). (a,d) Fixed grid at $512^2$, (b,e) co-scaling grid at $128^2$, (c,f) co-scaling grid at $512^2$. Color lines indicate $20$ equi-distant contours in magnetic pressure between $0$ and $3.7$. Streamlines have been overplotted in black. }
  \label{fig:blast3d}
\end{figure*}

At $t=0.15$, the domain has expanded to $[-1.418,1.418]^3$. We compare the result to a model run on a fixed grid of $512^3$ at this domain size. Figure~\ref{fig:blast3d} shows the mid-plane of the magnetic pressure in the $(x,y)$-plane (a) and in the $(x,z)$-plane (d) for the fixed-grid model. The co-scaling model at lower resolution shows the same morphology (b,e), though, because of fewer cells, the shock fronts are not as crisp, leading to a slightly larger appearance of the sphere. The co-scaling grid at $512^3$ reproduces the fixed-grid solution (c,f).

\subsection{A Comment on 3D Spherical Polar Coordinates}
We implemented and tested the co-scaling grid also for spherical-polar coordinates $(r,\theta,\phi)$. Spherical blast wave results are improved over fixed-grid models. Yet, for a uniform magnetic field on a full sphere, the co-scaling grid implementation suffers from the same limitations as documented for the stock version of Athena++\footnote{\href{https://github.com/PrincetonUniversity/athena-public-version/issues/3}{https://github.com/PrincetonUniversity/athena-public-version/issues/3}}. The resulting artifacts due to the singularity in the EMF appearing at the poles are somewhat reduced for the co-scaling grid but they still appear.

\section{Conclusion}
We extended the co-scaling grid method implemented in Athena++ \citep{HabeggerHeitsch2021} to MHD. We illustrate and detail the two corrections necessary for updating the magnetic field on a co-scaling grid: (1) Changes to the magnetic flux resulting from the change in length of the cell area boundary and (2) changes to the magnetic flux resulting from the change in cell area. We implement these adjustments in the Athena++ code, following the 2nd-order accurate, upwind constrained transport method introduced by \citep{GardinerStone2005}. 

We tested our implementation with one-, two-, and three-dimensional simulations. All tests showed either agreement with or improvement upon equivalent fixed grid simulations. The tests illustrate not only the accuracy of our implementation, but also the usefulness of the co-scaling grid method. Since all modifications are ``minimally invasive'' in the sense that they exploit the available Athena++ infrastructure, the MHD implementation performs and scales with processor number as discussed in HH21. For astrophysical fluid dynamics problems that involve drastic scale changes, the co-scaling grid will produce efficient and accurate simulations.

%%%%%%%%%%%%%%%%%%%%%%%%%%%%%
%%%%%%%%  Ack and Bib %%%%%%%
%%%%%%%%%%%%%%%%%%%%%%%%%%%%%
 
\section{Acknowledgments}
We thank the anonymous referee for thorough reports and detailed feedback. Computational resources provided by the University of North Carolina at Chapel Hill's Information Technology Services are gratefully acknowledged. FH is grateful for summer support from the School of Civic Life and Leadership (SCiLL) at UNC-CH. RH thanks NASA (NASA-FINESST Grant 80NSSC22K1749) for funding provided while this work was performed. 

\bibliography{sample631}{}

%%%%%%%%%%%%%%%%%%%%%%%%%%%%%
%%%%%%%%  Appendix    %%%%%%%
%%%%%%%%%%%%%%%%%%%%%%%%%%%%%

\appendix
\section{Areas and Reduced Dimensionality in Spherical Coordinates}
The area elements along the three coordinate directions $r,\theta,\phi$ are 
\begin{eqnarray}
  A^1_{i-\frac{1}{2},j,k} &\equiv& r^2_{i-\frac{1}{2}}\,\sin\theta_j\,d\theta\,d\phi\nonumber\\
  &=&r_{i-\frac{1}{2}}^2\,d(-\cos\theta_j)\,d\phi\\
  A^2_{i,j-\frac{1}{2},k} &\equiv& r_i\,dr\,\sin\theta_{j-\frac{1}{2}}\,d\phi\nonumber\\
  &=&d\left(\frac{r^2}{2}\right)_i\,\sin\theta_{j-\frac{1}{2}}\,d\phi\\
  A^3_{i,j,k-\frac{1}{2}} &\equiv& r_i\,dr\,d\theta = d\left(\frac{r^2}{2}\right)_i\,d\theta.
\end{eqnarray}
Note that the differential of the radial expression for $A^2$ and $A^3$ is not located at the cell center but at a volume-averaged position.
In the current implementation, we do not allow grid motion along $\theta$ or $\phi$, to keep the cell aspect ratios constant. Therefore, the time derivatives of the area expressions gain only contributions from the radial coordinate,
\begin{eqnarray}
  \frac{d}{dt}A^1_{i-\frac{1}{2},j,k} &=& 2r_{i-\frac{1}{2}}\,w^1_{i-\frac{1}{2}}\,d(-\cos\theta_j)\,d\phi \\
  \frac{d}{dt}A^2_{i,j-\frac{1}{2},k} &=& \left(r_{i+\frac{1}{2}}w^1_{i+\frac{1}{2}}-r_{i-\frac{1}{2}} w^1_{i-\frac{1}{2}}\right)\,\sin\theta\,d\phi\label{eqn:a2sph}\\
  \frac{d}{dt} A^3_{i,j,k-\frac{1}{2}} &=& \left(r_{i+\frac{1}{2}}w^1_{i+\frac{1}{2}}-r_{i-\frac{1}{2}} w^1_{i-\frac{1}{2}}\right)\,d\theta,\label{eqn:a3sph}
\end{eqnarray}
where $w^1_{i-\frac{1}{2}}$ is the wall velocity at position $r_{i-\frac{1}{2}}$. Note that the radial differentials for $A^2$ and $A^3$ (Equations~\ref{eqn:a2sph}, \ref{eqn:a3sph}) are discrete.

For simulations using Cartesian coordinates in two dimensions, the inactive dimension does not contribute to the apparent EMF (equation~\ref{e:apparentemf}), because of the "cell" length effectively being zero along the third dimension. For spherical coordinates, the situation is different, since cell extent along the inactive dimensions carries a factor of the radius.  While the $\theta$ and $\phi$ grids stay constant, the physical length scales $rd\theta$ and $r\sin\theta d\phi$ used to calculate the EMF do not. Therefore, {\em truly two-dimensional models in $(r,\theta)$ require updating all three components ($r$,$\theta$,$\phi$) of the EMF}.

\section{Total Variation in the Co-scaling Grid}\label{app:tvdcheck}
The total variation for single variable in a conservation law of the form of Equation~\ref{eqn:PDE} 
\begin{equation}
  TV(u^n) = \int|\partial_x u^n|dx\approx \sum_i |u^n_{i+1}-u^n_{i}|\label{eqn:tvd}
\end{equation}
measures the growth of spurious oscillations. In Equation~\ref{eqn:tvd}, $n$ is the time index and $i$ is the spatial index. The TV could be seen as a line integral along the 1D profile of the variable. If structure increases (e.g. shocks form) beyond the initial conditions, the time variation TV$(u^n)$ will increase. A scheme is called time-variation-diminishing (TVD) if TV$(u^{n+1})\leq$TV$(u^n)\,\forall n$. Though the Euler equations themselves do not obey the TVD property \citep{Toth2023}, the underlying advection equations for the characteristics should. A less stringent and more practical condition for solver stability introduced by \citet{Toth2023} requires that the Total of Time Variation (TOTV) 
\begin{equation}
  {\rm TOTV}(u^n) = \int |\partial_t u|dV \approx \sum_i\left|\frac{u_i^{n+1}-u_i^n}{t^{n+1}-t^n}\right|\,\Delta x^n_i\label{eqn:totv}
\end{equation}
is diminishing over time, i.e. the scheme is total-of-time-variation-diminishing (TOTVD) if TOTV$(u^{n+1})\leq$TOTV$(u^n)$. Note that Equation~\ref{eqn:totv} contains the spatially and time-dependent volume (here in 1D, the length) element $\Delta x_i^n$.

\begin{figure*}
  \includegraphics[width=\columnwidth]{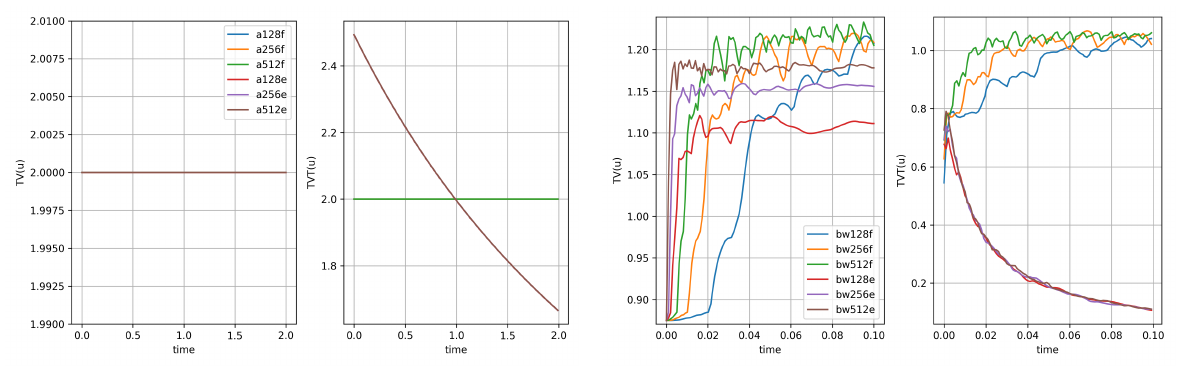}
  \caption{\label{fig:tvdcheck}TV (first and third panel) and TOTV (second and fourth panel) measured in the density for the advection test (see text) and the Brio-Wu test (Sec.~\ref{subsec:briowu}). Models with a trailing $f$ use a fixed grid, and models with a trailing $e$ use the co-scaling grid. Numbers indicate numerical resolution. Both fixed and expanding grid are TVD and TOTVD. For the Brio-Wu test, TVD is not expected initially, since the initial conditions lead to shocks and thus additional structure.}
\end{figure*}

Figure~\ref{fig:tvdcheck} summarizes the measurements of TV and TOTV of Athena++ for two one-dimensional tests. The left two panels show the TV and the TOTV for the advection of a density profile, while the two panels on the right show the same quantities for the Brio-Wu test (Sec.~\ref{subsec:briowu}). For the advection test, we define a tophat function in the density at constant pressure $p=1$ such that 
\begin{equation}
  \rho(x) = 1+\frac{1}{4}\left[1+\tanh\left(100\left(x+\frac{3}{2}\right)\right)\right]\left[1-\tanh\left(100\left(x+\frac{1}{2}\right)\right)\right]
\end{equation}
on a domain $-2\leq x \leq 2$. The profile is advected at a constant velocity of $v=1$ for a total time of $t=2$, i.e. at the end of the test the profile is centered on $x=1$. Boundary conditions are set to inflow ($\rho=1,p=1,v=1$) on the left and outflow on the right. For the expanding grid case, we stretch the grid at a constant rate such that at $t=2$, the grid has a size of $-3\leq x\leq 3$.

Both the fixed and the co-scaling grid are TVD and TOTVD for the advection test, as expected. For the Brio-Wu test, the initial conditions introduce shocks and thus additional structure, hence the TV levels out only after an initial increase. The expanding grid leads to an earlier saturation (leveling) of the TV, while the TOTV decreases monotonically. We conclude that the co-scaling grid implementation not only preserves the TVD and TOTVD properties of Athena++, but improves them.

\end{document}